# Gd disilicide nanowires attached to Si(111) steps

J. L. McChesney, A. Kirakosian, R. Bennewitz[a], J. N. Crain, J.-L. Lin, and F. J. Himpsel

Department of Physics, UW-Madison, 1150 University Ave., Madison, WI 53706, USA

[a] Dept. of Physics and Astronomy, University of Basel, 4056 Basel, Switzerland

### Abstract

Self-assembled electronic devices, such as quantum dots or switchable molecules, need self-assembled nanowires as connections. We explore the growth of conducting Gd disilicide nanowires at step arrays on Si(111). Atomically smooth wires with large aspect ratios are formed at low coverage and high growth rate (length >1µm, width 10nm, height 0.6nm). They grow parallel to the steps in the $[1\ \bar{1}\ 0]$ direction, which is consistent with a lattice match of 0.8% with the a-axis of the hexagonal silicide, together with a large mismatch in all other directions. This mechanism is similar to that observed previously on Si(100). In contrast to Si(100), the wires are always attached to step edges on Si(111) and can thus be grown selectively on regular step arrays.

With the advent of molecular electronics using self-assembled nanodots and nanotubes as electronic components there is a growing need for self-assembled interconnects. So far, electron beam lithography has been used for wiring individual devices, such as a memory cell or a transistor. This technique is not viable for assembling such devices by the billions. Therefore, a variety of methods have been studied for the directed self-assembly of wires between electrodes, for example selective bonding of a long DNA strand to short, complementary pieces of DNA at the electrodes and subsequent metallization [1].

Silicon surfaces are the traditional substrate for electronic devices, which makes them prime candidates for growing nanowires. It was found early on that silver nanowires with high aspect ratio can be obtained at Si(100) surfaces [2]. The explanation for such anisotropic growth behavior was a minimization of the strain energy by relaxing the strain perpendicular to a narrow wire while keeping it fully-strained in the long direction. More recently, transition metals [3-6] and rare earth [7-11] disilicides have attracted attention because of their anisotropic growth mode on Si(100). In that case the strain is anisotropic to begin with, due to good lattice match parallel to the growth direction combined with a mismatch perpendicular to it. For example, the hexagonal, Si-deficient $GdSi_{1.65}$ phase has a lattice match of 0.8% between its a-axis and the $[1\ \bar{1}\ 0]$ direction of Si while being mismatched by 10-20% in other orientations [12,13]. Since $[1\ \bar{1}\ 0]$ is an in-plane direction common to both Si(100) and Si(111), there is the possibility of growing nanowires on both surfaces. Further benefits of these silicides are their conductivity (e.g., 190 µΩcm for hexagonal $GdSi_{1.65}$ at room temperature [13]) and their extremely low Schottky barrier to n-type silicon [14]. Another feature of interest is their magnetism (anitferromagnetic below $T_N = 33$ K).

A primary requirement for the use of nanowires in self-assembled devices is the ability to control their location. If the active electronic elements, such as memory cells and switches, are assembled first, then the wiring has to find their location and start the growth there. Alternatively, a wire matrix can be assembled first, and then the switching components inserted at the cross-links [15,16]. Previous work with rare earth disilicide nanowires has focused on the Si(100) orientation, where two sets of orthogonal nanowires grow at the two types of terraces that are rotated by $90^0$. However, the wires apparently nucleate at random locations on the surface [7-11]. Therefore, we have turned our attention to Si(111), where it is possible to produce highly-regular arrays of steps [17] that might serve as templates for forming wire arrays. Indeed, we find that Gd disilicide wires exclusively grow along step edges on Si(111), as shown in Figs. 1 and 2. That suggests their possible use in crossed-wire arrays with back-to-back wafers bonded by the molecules or nanodots.

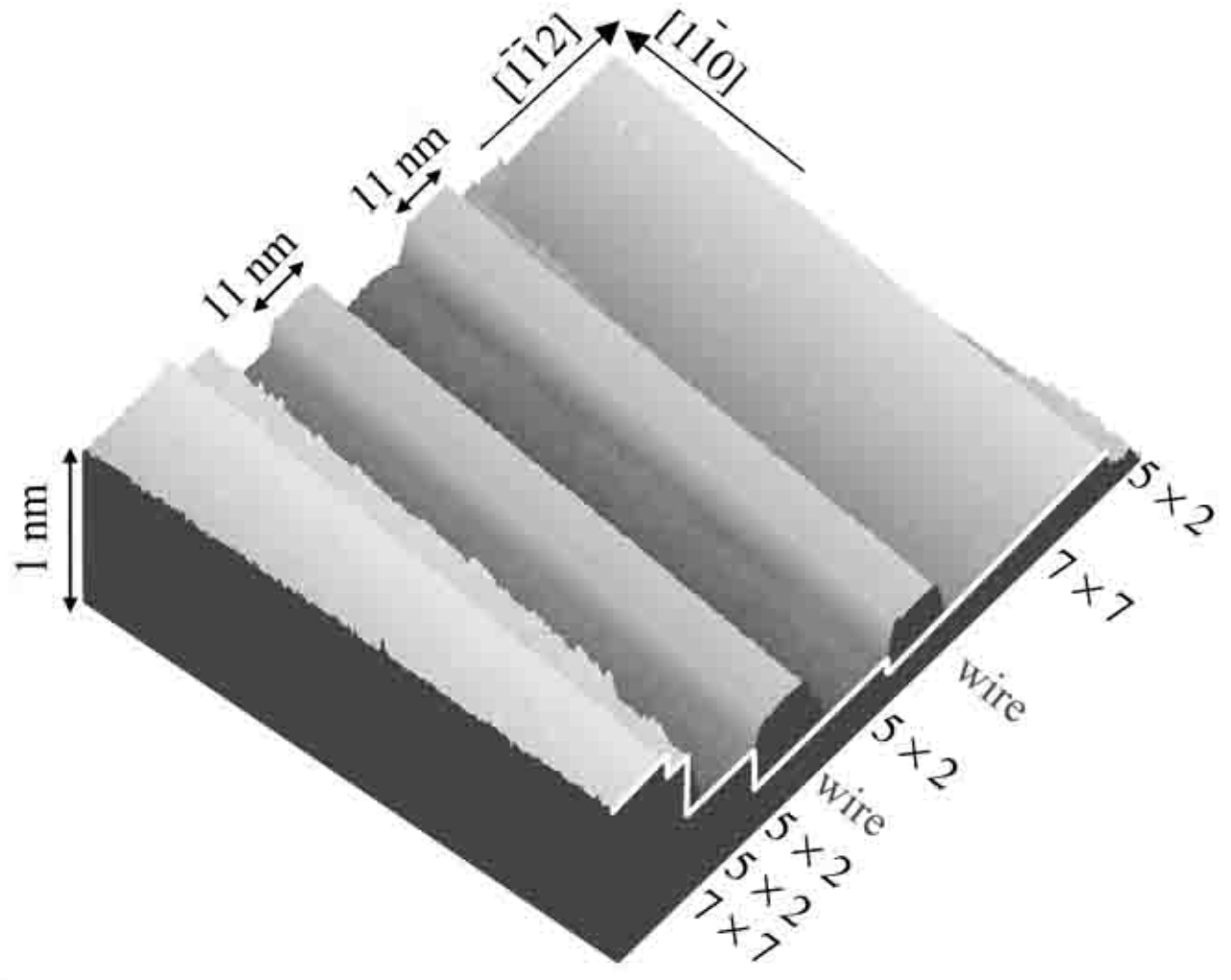

**Fig. 1** Two Gd disilicide nanowires attached to step edges of Si(111). Note the atomically-precise widths of 11 nm. (STM image at +1.2V sample bias).

Vicinal Si (111) with a 1° miscut toward the $[\bar{1}\bar{1}2]$ azimuth serves as a template since the $[1\bar{1}0]$ step direction has an 0.8% lattice match with the a-axis of bulk hexagonal silicide. Perpendicular to the step (along the $[\bar{1}\bar{1}2]$ direction) there is a large mismatch, no matter which orientation the silicide takes relative to the Si substrate. If the c-axis lies perpendicular to the surface, as reported in [12] for silicide films, the mismatch is 9%. If the c-axis is in-plane the mismatch is 28%. For optimal growth of long wires distributed evenly across the surface, 0.05 monolayers of Gd where deposited at a rate of 1.3 Å/min while keeping the sample at 675°C. Although other temperatures and flux rates produced wires, these parameters produced the longest and most densely spaced wires.

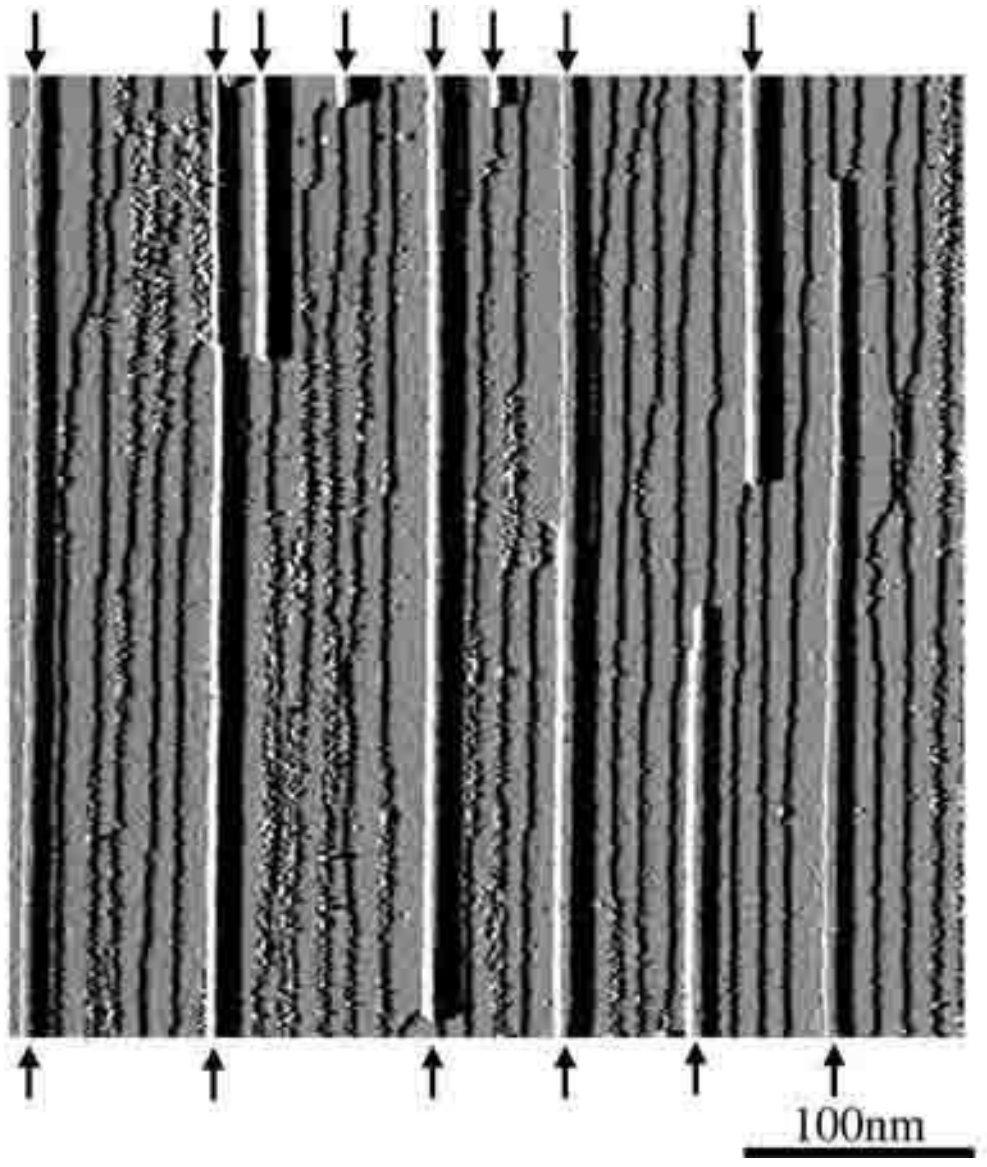

**Fig.2** STM image of several Gd disilicide nanowires (arrows) that are all attached to step edges of Si(111) running along the $[1\bar{1}0]$ direction. The x-derivative of the topography is shown in this and the following images. Wires appear to be illuminated from the left and casting shadows to the right, which is downhill.

There is a very limited window within which wires will be produced due to the kinetic nature of the process. The flux rate needs to exceed the diffusion rate of Gd into bulk Si, which is significant at 700°C [18]. Furthermore, the Gd-induced Si(111)5×2-Gd surface structure [19] competes with nanowires for available Gd. Though Si(111)5×2-Gd can coexist with nanowires on the same surface (see Fig. 1), the highest wire densities were observed when the remaining surface was still clean Si(111)7×7. There is a second Gd-induced structure at higher coverage than 5×2-Gd, i.e. Si(111)√3×√3-Gd. It is able to coexist with nanowires as well.

Typical dimensions of the nanowires are a length of >1μm, a width of 10nm, and a height of 0.6nm. Figure 3 shows wires in the μm range, whose length is mainly limited by kinks of the underlying Si(111) steps. In fact, the wires straighten out the steps to atomic smoothness until a step deviates so far from a straight line that it requires too much Si needs to be displaced to straighten out the step. This is reminiscent of the growth mode on the Si(100) surface, where a nanowire can grow only on the same terrace. It extends an atomically flat terrace as far as feasible within the limited Si diffusion rate by building up material on a downhill terrace or burrowing into an uphill terrace. Such an effect is also seen on Si(111).

The width of the nanowires remains atomically precise along the wire in most cases (see Figs. 1-3). The most frequent width is 11 nm (Fig. 1), but widths between 5 and 20 nm have been observed. Occasional changes in the wire width tend to occur in multiples of 0.4 nm, which would indicate that this is the maximum width for dislocation-free epitaxy perpendicular to the wires.

Strain confines not only the wire width, but the height as well. All wires are 0.6 nm high relative to the bottom terrace, independent of wire density or deposition time. Only slight variations are seen (0.05nm), which can be traced to different substrate reconstructions and tips with differences in the electronic structure.

The Gd disilicide nanowires are rather inert and stable. After being exposed to residual gas at 1×10$^{-10}$ Torr for several days, the wires are unaffected while the 5×2-Gd structure becomes completely covered with adsorbates (grainy patches in Fig. 2). After several 10 min anneals at 700°C the wires appear unchanged but the surrounding 5×2-Gd surface converts back to a mixture of clean 7×7 and patches of multi-domain 5×2-Gd.

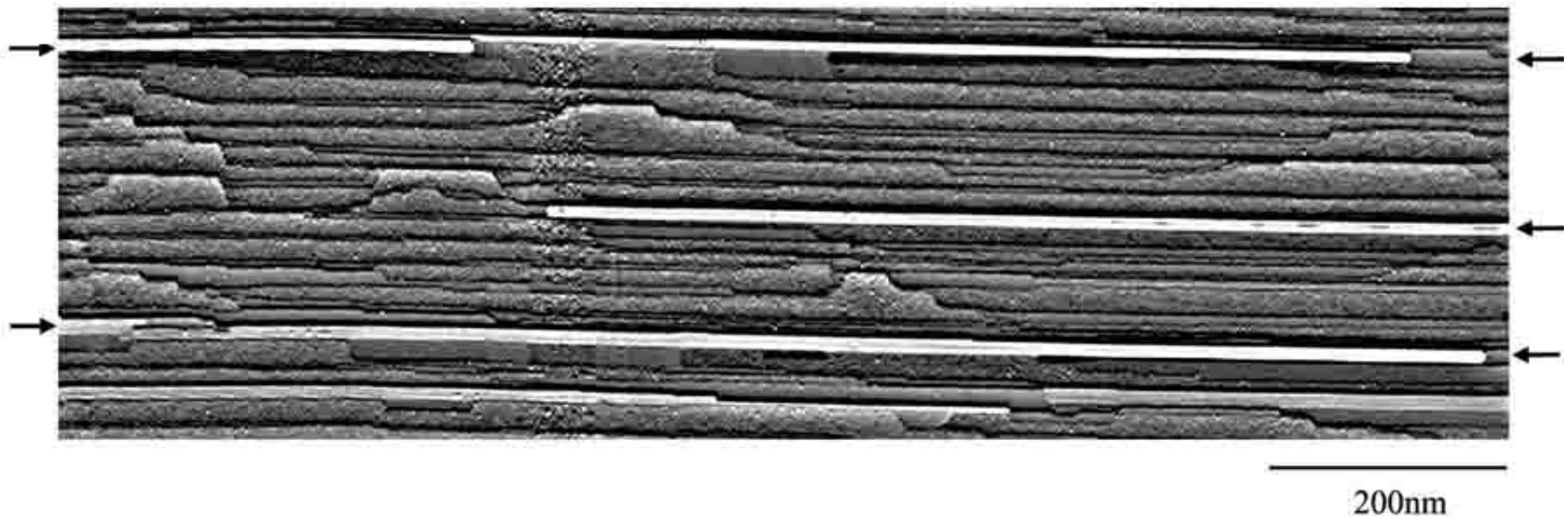

**Fig. 3** Nanowires reaching μm dimensions. Their length is limited by the straightness of the Si steps, because the wires do not form kinks.

The surface of the wires exhibits chain like surface reconstructions, one with a chain spacing of two silicon surface lattice constants, the other with four. Both are shown Fig. 4, where a nanowire is sandwiched between a 5×2-Gd terrace on the left and a 7×7 terrace on the right. There is a significant change when tunneling from occupied orbitals (−2V sample bias) to tunneling into unoccupied orbitals (+2V bias), which hints at a pronounced modulation of the charge distribution perpendicular to the chains, similar to that seen for the 5×2-Gd structure [19].

In summary, we have observed spontaneous growth of atomically-precise Gd disilicide nanowires on Si(111). The typical width is 10nm and the length can exceed a μm. The growth mechanism is driven by lattice match between the a-axis of the silicide and the $[1\bar{1}0]$ direction along the Si surface, as observed for a number of rare earth disilicides on Si(100). However, on Si(111) the nanowires grow only at step edges. That allows some degree of control over their position, which is the first step towards self-assembled interconnects between nano-devices at a surface. The next step will be the search for a nucleation mechanism that would make it possible to position the beginning or end of a nanowire along a step.

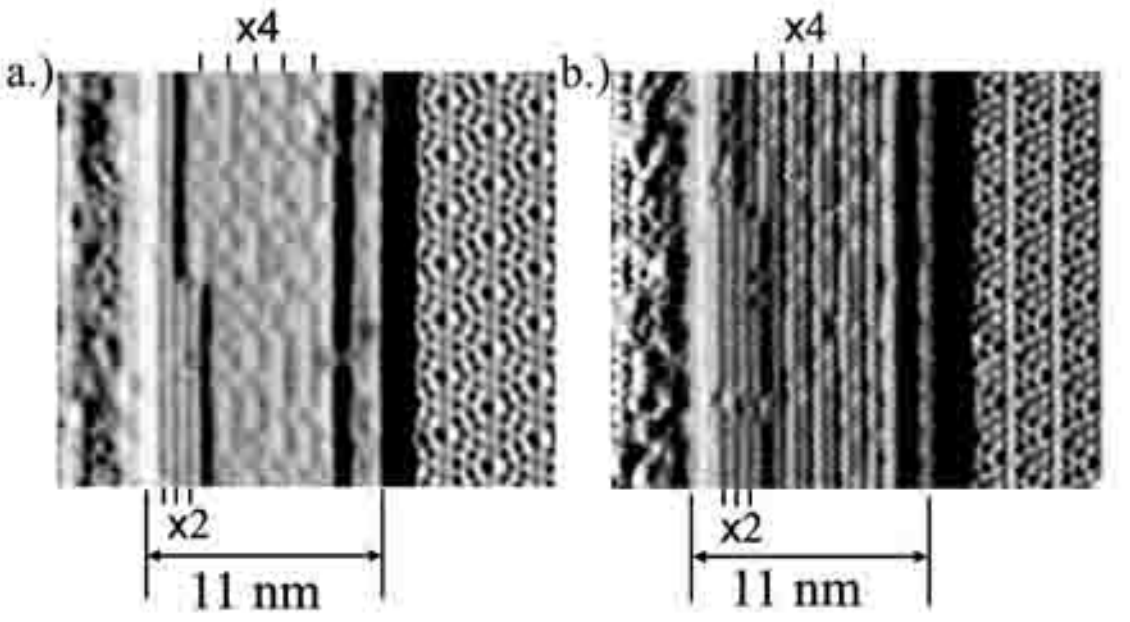

**Fig. 4** Close-up of the surface reconstruction on top of a nanowire at two sample bias voltages (+2V on the left, −2V on the right). Two types of chain structures are observed with a chain spacing of 2 and 4 lattice constants (×2 and ×4).

Acknowledgment: This work was supported by the NSF (DMR-9815416 and DMR-0079983) and by the DOE under Contract No. DE-FG02-01ER45917.